\documentclass[11pt,a4paper]{article}
\usepackage[T1]{fontenc}
\usepackage[utf8]{inputenc}
\usepackage{lmodern,amssymb}
\usepackage[margin=26mm,headheight=14pt]{geometry}
\usepackage{microtype}
\usepackage{amsmath,amsthm,mathtools}
\usepackage{todonotes}
\usepackage[
backend=biber,
style=alphabetic,
citestyle=alphabetic,
maxalphanames=6,
maxcitenames=99,
mincitenames=98,
maxbibnames=99,
giveninits=true,
url=false,
doi=true,
isbn=true,
backref=true
]{biblatex}
\DeclareNumChars*{:}

\usepackage{algorithm}
\usepackage{algpseudocode}

\usepackage{xfrac}
\usepackage{booktabs,tabularx,longtable,array}
\usepackage{enumitem}
\usepackage{float,flafter}
\newfloat{algorithm}{tbp}{loa}
\floatname{algorithm}{Algorithm}
\usepackage{fancyhdr}
\usepackage[hidelinks]{hyperref}
\usepackage[capitalize,nameinlink,noabbrev]{cleveref}
\crefname{algorithm}{Algorithm}{Algorithms}
\crefname{theorem}{Theorem}{Theorems}
\Crefname{lemma}{Lemma}{Lemmas}
\Crefname{corollary}{Corollary}{Corollaries}
\Crefname{claim}{Claim}{Claims}
\Crefname{fact}{Fact}{Facts}
\Crefname{remark}{Remark}{Remarks}
\Crefname{observation}{Observation}{Observations}
\Crefname{figure}{Figure}{Figures}
\Crefname{line}{Line}{Lines}
\Crefname{stepsromani}{Step}{Steps}
\Crefname{stepsarabici}{Step}{Steps}
\Crefname{stepsalph}{Step}{Steps}

\usepackage{needspace,aliascnt}
\newtheorem{theorem}{Theorem}[section]
\newaliascnt{lemma}{theorem}
\newtheorem{lemma}[lemma]{Lemma}
\aliascntresetthe{lemma}
\newaliascnt{observation}{theorem}
\newtheorem{observation}[observation]{Observation}
\aliascntresetthe{observation}
\newaliascnt{proposition}{theorem}

\aliascntresetthe{proposition}
\newaliascnt{corollary}{theorem}

\aliascntresetthe{corollary}
\theoremstyle{definition}
\newaliascnt{definition}{theorem}
\newtheorem{definition}[definition]{Definition}
\aliascntresetthe{definition}
\newaliascnt{example}{theorem}

\aliascntresetthe{example}
\newaliascnt{remark}{theorem}

\aliascntresetthe{remark}
\numberwithin{equation}{section}
\DeclareMathOperator{\Ball}{Ball}

\newcommand{\eps}{\varepsilon}
\renewcommand{\epsilon}{\varepsilon}
\DeclareMathOperator{\supp}{supp}
\newcommand{\OPT}{\mathrm{OPT}}

\newcommand{\ceil}[1]{\left\lceil#1\right\rceil}

\newcommand{\poly}{\operatorname{poly}}

\newcommand{\RMFC}{\textnormal{RMFC}}
\renewcommand{\P}{\mathsf{P}}
\newcommand{\NP}{\mathsf{NP}}

\definecolor{ETHBlue}{HTML}{215CAF}
\definecolor{ETHDark}{HTML}{1F2937}
\definecolor{ETHLight}{HTML}{EAF2F8}
\definecolor{ETHGreen}{HTML}{4E9F3D}
\definecolor{ETHOrange}{HTML}{F2A541}
\definecolor{ETHRed}{HTML}{C0392B}
\definecolor{ETHGray}{HTML}{6B7280}

\usetikzlibrary{arrows.meta,positioning,calc,decorations.pathreplacing}

\setlist{topsep=4pt,itemsep=2pt,parsep=0pt}
\title{Nearly Optimal Dilation for Non-Uniform \texorpdfstring{$k$}{k}-Center with Constant-Factor Budget Violation}
\author{Jannis Blauth\thanks{
		Department of Mathematics, ETH Zurich, Zurich, Switzerland.
		Email: \href{mailto:jblauth@ethz.ch}{jblauth@ethz.ch}.}
	\and
	Siri Prüßner\thanks{
		Department of Mathematics, ETH Zurich, Zurich, Switzerland.
		Email: \href{mailto:spruessner@ethz.ch}{spruessner@ethz.ch}.}
}
\date{}

\begin{document}
\maketitle
\begin{abstract}
In Non-Uniform $k$-Center, we are given a metric space $(V,d)$ and $L$ levels of balls with prescribed radii $r_1 \ge \ldots \ge r_L$ and nonnegative integer budgets $k_1,\ldots,k_L$.
The goal is to place at most $k_\ell$ centers of each level $\ell \in [L]$ in $V$ and minimize the dilation $\lambda$ such that the corresponding balls of radii $\lambda r_1,\dots,\lambda r_L$ cover $V$.
While this problem admits no subexponential approximation factor unless $\P=\NP$ \cite{NUkC}, the picture changes drastically when additional centers are allowed.
Following the definition in \cite{blauth2025},
an $(\alpha,\beta)$-bicriteria approximation to Smooth Non-Uniform $k$-Center covers $V$ with dilation $\beta \lambda_{\textnormal{opt}}$ such that, for each $\ell \in [L]$, the number of centers of the first $\ell$ levels is at most $\alpha$ times their total budget.

For every fixed $\varepsilon>0$, we give a randomized $(1+\varepsilon,8+\varepsilon)$-approximation and a deterministic $(\mathcal{O}_{\varepsilon}(1),2+\varepsilon)$-approximation, both running in polynomial time.
Even for classical $k$-Center, obtaining an $(\mathcal{O}(1),2-\delta)$-approximation is NP-hard for every fixed $\delta \in (0,1]$.
Thus, our second result almost settles the optimal dilation when a constant-factor violation of the budget is allowed, unless $\P=\NP$.

We combine a new primal-dual enumeration procedure with a simple bucketing argument that yields a new trade-off between the two approximation factors.
It remains open whether there is a polynomial-time
$(1+\varepsilon,2+\varepsilon)$-bicriteria approximation,
which would almost match the classical $2$-approximation for $k$-Center \cite{HochbaumShmoys1985KCenter}.
\end{abstract}

\section{Introduction}\label{sec:intro}

The metric $k$-Center problem asks for a collection of $k$ balls of minimum common radius which covers a given finite metric space. A natural generalization, introduced by \textcite{NUkC}, is the \emph{non-uniform $k$-center problem} (NUkC), in which balls of several different radii are available. 
More precisely, we are given a finite metric space $(V,d)$
with $n:=|V|$, radii $r_1\ge \ldots\ge r_L\ge 0$, and budgets $k_1,\dots,k_L \in \mathbb{Z}_{\ge 0}$. 
We call $(V,d,k,r)$ an instance with $L$ levels.
The objective is to find the smallest $\lambda\ge0$ such that $V$ can be covered by opening, for each $\ell\in[L]$, at most $k_\ell$ balls of level $\ell$, each with radius $\lambda r_\ell$ and center in $V$.

A motivating interpretation, already discussed in \cite{NUkC}, is the placement of emergency vehicles with different speeds. Suppose that $k_\ell$ vehicles of speed $r_\ell$ are available for each $\ell\in[L]$. We wish to choose their locations in $V$ so as to minimize the maximum time needed to reach a request at any point of $V$. A vehicle of speed $r_\ell$ can reach precisely the points within distance $T r_\ell$ of its location in time $T$. Thus the smallest achievable worst-case response time is exactly the optimum dilation of the corresponding NUkC instance. Increasing the budgets allows additional vehicles of each type, while increasing the dilation allows a longer response time.

Allowing different radii makes the problem considerably harder than regular k-Center. \textcite{NUkC} proved that approximating the optimum dilation within any factor $\beta \le 2^{\poly n}$ is NP-hard, even on tree metrics. On the other hand, they showed that constant-factor guarantees become possible if one allows a controlled violation of the budgets. Therefore, it makes sense to consider bi-criteria approximations: An $(\alpha,\beta)$-approximation may use, for each $\ell \in [L]$, at most $\alpha k_\ell$ balls of level $\ell$ and radius $\beta \lambda_{\textnormal{opt}} r_\ell$. 
Here, $\lambda_{\textnormal{opt}}$ denotes the optimal dilation subject to the original budget constraints, while $\beta$ denotes the algorithm’s multiplicative approximation factor for the dilation.
Their algorithm achieves a $(24,10)$-approximation. Subsequently, \textcite{blauth2025} obtained a deterministic $(2,15+\varepsilon)$-approximation for every fixed $\varepsilon>0$.
Note that it is NP-hard to compute a $(2-\delta, \mathcal{O}(1))$-approximation for any $\delta \in (0,1]$. Indeed, by replacing the instance by an equivalent one with $k_\ell = 1$ for each $\ell \in [L]$, opening one additional center at any level already doubles that level’s budget.

However, the guarantee in \cite{blauth2025} is even stronger in the following sense. 
An $(\alpha,\beta)$-approximation for \emph{Smooth Non-Uniform $k$-Center} (SNUkC) is a feasible solution with balls
of radius $\beta\lambda_{\textnormal{opt}} r_\ell$ for each level $\ell$, and using at most
$\alpha k_{\le \ell}$ balls from the first $\ell$ levels for each $\ell\in[L]$, where $k_{\le \ell} := \sum_{i=1}^\ell k_i$ is the cumulative budget up to level $\ell$.
In this framework, their algorithm computes a $(1+\eps,15+\eps)$-approximation.
In words, for each $\ell \in [L]$, the number of centers of the $\ell$ levels with the largest radii may exceed their total budget by a factor of at most $1+\eps$. 
Thus, allowing an arbitrarily small violation of the budget constraints yields a constant-factor approximation, whereas strict adherence to these constraints makes it NP-hard to achieve any subexponential approximation factor.
In this paper, we study the dilation achievable under both arbitrarily small and constant-factor budget violations.

\subsection{Main results}

First of all, we present a novel randomized enumeration approach that beats the previous best-known dilation of $15+\eps$ if one only allows an arbitrarily small violation of the budget constraints \cite{blauth2025}.
Here, randomization means that for every fixed input, the probability that it achieves the stated approximation guarantee can be made arbitrarily close to one.

\begin{theorem}\label{thm:main_rand}
	For every fixed $\eps>0$, there is a randomized polynomial-time $\left(1+\eps,8+\eps\right)$-approximation algorithm for SNUkC.
\end{theorem}

We complement this result by providing an almost optimal approximation of the dilation. 
First, we show a dilation factor arbitrarily close to two, at the expense of increasing the resource factor. 
Here, $\mathcal{O}_\eps(1)$ denotes a constant that depends on $\eps$.

\begin{theorem}\label{thm:main}
	For every fixed $\eps>0$, there is a deterministic polynomial-time $\left(\mathcal{O}_\eps(1),2+\eps\right)$-approximation algorithm for SNUkC.
\end{theorem}

We discuss how to obtain both theorems in \cref{sec:main-results-formal} and provide a formal proof in \cref{sec:parm_readjustment} for completeness.

The tight inapproximability result for Dominating Set implied by \textcite{DS14} readily yields the following hardness result for SNUkC that even holds for classical $k$-Center. In particular, achieving a dilation factor below two requires a logarithmic violation of the budget constraints, unless $\P  = \NP$. 
For completeness, we provide a short proof in \cref{sec:hardness}.

\begin{observation}\label{thm:hardness}
For every fixed $0<\eps<1$, it is NP-hard to compute a $((1-\eps)\ln n,2-\eps)$-approximation for SNUkC, even when there is only one radius.
\end{observation}

Note that our approximation guarantee establishes a trade-off between the two approximation factors. 
It remains open whether one can simultaneously achieve the best possible approximation factors for both the budget violation and the dilation of the radii, namely a $(1+\eps,2+\eps)$-approximation for SNUkC, almost closing the gap to the approximability of classical $k$-Center.

\subsection{Techniques and proof ideas}

At a high level, our algorithm follows a similar approach as in \cite{NUkC, blauth2025}.
\textcite{NUkC} observed that if the budget of the largest radius is sufficiently large, one can round any fractional solution to a standard linear programming relaxation while losing a small factor on the budget violation and a factor of $8$ on the dilation. 
This LP-rounding approach leverages their LP-aware reduction to \emph{Resource Minimization for Fire Containment on Trees} (RMFC-T), as well as an extreme-point rounding argument for RMFC-T by \textcite{10.1145/3173046}.
We comment on the relation between SNUkC and RMFC-T in \cref{sec:rounding}.
By using a simple bucketing idea,
we introduce a new trade-off between the two approximation factors when rounding fractional solutions.
This allows us to reduce the loss in the dilation of the radii to a factor arbitrarily close to two.

However, there is no guarantee that the budget of the first radius is sufficiently large, as required to bound the violation of the budget constraint.
Instead, we use a known compression technique introduced for RMFC-T by \textcite{10.1145/3173046} and adapted to NUkC in \cite{NUkC,blauth2025}.
This guarantees that the number of different
radii is logarithmic in the number of vertices, and the cumulative budgets
are geometrically increasing.
This incurs an arbitrarily small multiplicative loss in both
the budget violation and the dilation.
The high-level idea is then to identify points close to centers with large radius in an optimum solution via an enumeration approach---and open centers of slightly increased radius at these points---such that the remaining vertices not covered by these balls can be fractionally covered by balls of small radius. 
For the fractional solution using only small radii, we can apply the LP rounding described above, thanks to the growth of the cumulative budgets after compression.

The enumeration approaches in \cite{NUkC} and \cite{blauth2025} achieve dilation factors of $10$ and $15+\eps$, respectively.
At a high level, \textcite{NUkC} select a small set of well-separated clients for each large radius and branch on whether an optimal ball of that radius is centered near a selected client.
\textcite{blauth2025} compute a sparse fractional cover and guess how
its large balls are positioned relative to optimal balls of different radii.
We provide a novel enumeration approach by using the dual of the covering LP to guide our choice of centers for large balls.
Our approach not only improves the dilation guarantee but also simplifies both the enumeration and its analysis.
In each step of the enumeration, we test whether the uncovered points admit a fractional cover using only balls of small radius, with a slight increase in their budgets. If the test fails, we compute an optimum dual solution that assigns nonnegative weights to the uncovered clients.
Along an enumeration branch consistent with a fixed optimum solution,
these weights have the property that a constant fraction of their total weight is covered by large balls for which we did not already choose a center.
Consequently, by sampling a client according to these weights, with constant probability, we sample a vertex $u$ that is covered by a large ball of radius $r$ in the fixed optimum solution for which we did not already choose a center. By the triangle inequality, opening a ball at $u$ of radius $2r$ covers the entire original ball of the optimum solution. This allows us to reduce the dilation to two for the center with large radius.
With slightly worse constants, this approach can be derandomized using a hitting set argument.

Once the remaining clients admit the required fractional cover by small balls, we apply the LP-rounding approach discussed above.

\subsection{Further related work}

Besides the mentioned bi-criteria approximations for (Smooth) Non-Uniform $k$-Center, there has been a long line of research studying the approximability of this problem if the number of different radii is constant. 
Note that for this special case, the approximation hardness by \textcite{NUkC} does not apply.
The following algorithms respect the budget for each radius type exactly;
their approximation guarantees therefore refer solely to the dilation of the radii.

\textcite{NUkC} presented a $(1+\sqrt 5)$-approximation for two different radii, which was later extended to a $22$-approximation for three different radii by \textcite{jiaNonUniformKCenterConstant2022}, and then to an $\mathcal{O}(1)$-approximation for four different radii by \textcite{inamdarNonUniformKCenterGreedy2022}.
Recently, \textcite{InamdarLochetPanolan2026NUkC} have found a constant-factor approximation algorithm for any fixed constant number of different radii.

A closely related problem is $k$-center with outliers, also called
robust $k$-center, where up to $\ell \in \mathbb{Z}_{\ge 0}$ points may remain uncovered.
This is a special case of NUkC with $k$ balls of a common positive
radius and $\ell$ balls of radius zero.
\textcite{charikarFacilityLocationOutliers2001} gave a
$3$-approximation for this problem, which was subsequently improved
to an optimal $2$-approximation unless $\P = \NP$ by \textcite{NUkC};
\textcite{harrisLotteryCenterOutliers2017} independently obtained
the same approximation guarantee.
For the non-uniform variant with two different radii and outliers,
\textcite{chakrabartyRobustTwoRadii2023} gave a
$10$-approximation using a round-or-cut framework.
\textcite{jiaNonUniformKCenterConstant2022} subsequently provided
a simpler analysis of this guarantee and gave a reduction
that transforms an $\alpha$-approximation for the robust variant
with $t-1$ different radii into a $(2\alpha+2)$-approximation
for NUkC with $t$ different radii.
Finally, \textcite{inamdarNonUniformKCenterGreedy2022} obtained an
$\mathcal{O}(1)$-approximation for the robust variant with three different
radii, which underlies their result for four radii without outliers.

NUkC has also been studied from the perspective of parameterized
approximation.
\textcite{bandyapadhyayParameterized2024} gave a
$2$-approximation with running time $2^{\mathcal{O}(k\log k)}n^2$,
where $k$ denotes the total number of balls.
For the more general variant in which candidate centers are
specified separately from the points to be covered, their
approximation guarantee is $3$.
For the continuous Euclidean variant, where centers may lie anywhere in $\mathbb{R}^d$, they obtain a $(1+\eps)$-approximation in time $2^{\mathcal{O}(k\log k/\eps)}nd$.
For shortest-path metrics of unweighted graphs,
\textcite{lieskovskyGraphBurning2022} show that NUkC can be solved
exactly in polynomial time when both the number of distinct radii
and the treewidth of the underlying graph are constant.

\subsection{Outline}

\cref{sec:model} introduces the model, states our precise approximation guarantees, and gives a high-level overview of our primal-dual guided enumeration algorithms and the underlying ideas. In \cref{sec:algo_analysis}, we formally prove the approximation guarantees and running time bounds of both the randomized and deterministic algorithms. \cref{sec:rounding} discusses the LP-aware reduction to Resource Minimization for Fire Containment and the resulting rounding procedure for fractional covers using our bucketing idea.

\section{High-level outline of our approach}\label{sec:model}

In this section, we formally introduce the model, discuss some well-established reductions, and formally state the guarantees of our algorithms, leading to our main results.
Along the way, we highlight the high-level ideas of our approaches.

\subsection{Reduction to compressed instances of Smooth NUkC}

Let $(V,d,k,r)$ be an instance of NUkC with $L$ levels.
For $v \in V$ and $r \in \mathbb{R}_{\ge 0}$, we denote by
\[
 \Ball(v,r)=\{u\in V:d(u,v)\le r\}
\]
the ball around $v$ of radius $r$.
For $A\subseteq V\times[L]$, let 
\[
 A_\ell= A \cap (V \times \{\ell\}),\qquad 
 A_{\le\ell}= \bigcup_{i=1}^\ell A_{i},\qquad \text{and} \qquad A_{>\ell}= A \setminus A_{\le \ell}.
\]
For $\alpha, \beta \in \mathbb{R}_{\ge 0}$, we call $A$ an \emph{$(\alpha,\beta)$-feasible solution to NUkC} if 
\begin{itemize}
	\item $|A_{\ell}|\le \alpha k_\ell$ for each $\ell \in [L]$, and
	\item $\bigcup_{(v,\ell) \in A} \Ball(v,\beta r_\ell) = V$.
\end{itemize}
In words, $A$ covers the whole metric space when dilating the balls by a factor of $\beta$, while exceeding the budget for each level by a factor of at most $\alpha$.

As mentioned earlier, it is NP-hard to compute an $(\alpha,\beta)$-approximation for any $\alpha < 2$ and $\beta \le 2^{\poly n}$ \cite{NUkC}.
This hardness follows directly from the hardness of computing a $(1,\beta)$-approximation for any $\beta \le 2^{\poly n}$, since any instance can be transformed into an equivalent instance with $k_\ell = 1$ for every $\ell \in [L]$.
This motivates the study of Smooth Non-Uniform $k$-Center (SNUkC), introduced by \cite{blauth2025}, which measures budget violations in a more continuous manner.
For each $\ell \in [L]$, let $k_{\le \ell} = \sum_{i=1}^{\ell} k_i$, and set $k_{\le 0}:=0$.

Let $\alpha,\beta \in \mathbb{R}_{\ge 0}$. 
An instance $(V,d,k,r)$ of SNUkC is defined as for NUkC, except that we do not require $k$ to be integral.
Given such an instance, we call a set $A \subseteq V \times [L]$ an \emph{$(\alpha,\beta)$-feasible solution to SNUkC} if
\begin{itemize}
	\item $|A_{\le \ell}| \le \alpha k_{\le \ell}$ for each $\ell \in [L]$, and
	\item $\bigcup_{(v,\ell) \in A} \Ball(v,\beta r_\ell) = V$.
\end{itemize}
 In other words, $A$ covers the entire metric space after dilating the balls by a factor of $\beta$, while exceeding the cumulative budget for the $\ell$ levels with the largest radii by at most a factor of $\alpha$, for every $\ell \in [L]$.
These cumulative constraints reflect the greater value of larger balls compared to smaller ones.

This framework allows constant-factor approximations with multiplicative budget violations arbitrarily close to one, while also simplifying the algorithms and their analysis.
When referring to numbered results in \cite{blauth2025}, we use the numbering of the full version \cite{blauth2025Full}.
The following result establishes a close relationship between NUkC and its smooth variant.

\begin{lemma}[Observation 6.1 in \cite{blauth2025Full}]
	Let $\alpha,\beta \in \mathbb{R}_{\ge 0}$.  An $(\alpha, \beta)$-approximation algorithm for SNUkC implies an $(\lceil \alpha \rceil, \beta)$-approximation algorithm for NUkC.
\end{lemma}

As mentioned earlier, the high-level idea of our algorithm is to combine an enumeration approach  with a rounding approach for fractional solutions.
For both subroutines, it will turn out to be beneficial to reduce the number of different radii.

\begin{definition}[Adapted from Definition~6.3 in \cite{blauth2025Full}]
	For $\epsilon > 0$, an SNUkC instance $(V,d,k,r)$ with $L$ levels and $n$ vertices is $\epsilon$-compressed if
	\begin{itemize}
		\item $L \leq \ceil{\log_{1+\varepsilon}(n)} + 1$,
		\item $k_1 \geq 1$ and $k_{\leq \ell+1} \geq (1+\varepsilon)k_{\leq \ell}$ for each $\ell \in [L-1]$, and
		\item $r_\ell \geq (1+\varepsilon)r_{\ell+1}$ for each $\ell \in [L-1]$.
	\end{itemize}
\end{definition}

In words, the number of levels is logarithmic in the number of vertices, the cumulative budgets increase geometrically, and the radii decrease geometrically.

The construction in \cite[Theorem~7.2]{blauth2025Full} yields the required geometric growth of cumulative budgets. It preserves $(1,1)$-feasibility and lifts solutions with a factor $1+\eps$ loss in each criterion.

For $\alpha,\beta \in \mathbb{R}_{\ge 0}$, we call an instance of SNUkC $(\alpha,\beta)$-feasible if it admits an $(\alpha,\beta)$-feasible solution. 
The following theorem shows that by guessing the optimum dilation, scaling the radii accordingly, and compressing the resulting instance, we can restrict our attention to $(1,1)$-feasible $\eps$-compressed instances of SNUkC.

\begin{theorem}[Theorem 6.4 in \cite{blauth2025Full}]\label{thm:reduction}
	Let $\varepsilon > 0$. An algorithm computing an $(\alpha,\beta)$-feasible solution for $(1,1)$-feasible $\varepsilon$-compressed SNUkC instances implies a $((1+\varepsilon)\alpha, (1+\varepsilon)\beta)$-approximation for (general) SNUkC instances.
\end{theorem}

\subsection{Our approximation guarantees}\label{sec:main-results-formal}

For $(1,1)$-feasible instances of $\eps$-compressed SNUkC, we get the following guarantees, which, together with \cref{thm:reduction}, directly imply our main results, namely \cref{thm:main_rand,thm:main}.
For completeness, we provide a formal proof of \cref{thm:main_rand,thm:main} in \cref{sec:parm_readjustment}.
For every positive integer $b$, let
\begin{equation*}
	\rho_b:=\frac{2(b+1)^{1+1/b}}{b}.
\end{equation*}

\begin{theorem}\label{thm:det}
	For every $b\in \mathbb{Z}_{>0}$ and every $\eps>0$, there
	is a deterministic polynomial-time algorithm that, given a $(1,1)$-feasible $\eps$-compressed instance of
	SNUkC, computes a $(2b + \eps, \rho_b )$-feasible solution.
\end{theorem}

If randomization is allowed, we can improve the budget violation from $2b+\eps$ to $b+\eps$, while keeping the same dilation.
As in \cref{thm:main_rand}, randomization means that for every fixed input, the probability that it achieves the stated approximation guarantee can be made arbitrarily close to one.

\begin{theorem}\label{thm:rand}
	For every $b\in \mathbb{Z}_{>0}$, $\eps>0$, and $p \in (0,1)$, there is a randomized polynomial-time algorithm that computes, for a given $(1,1)$-feasible $\varepsilon$-compressed instance of SNUkC, a $(b+\eps,\rho_b)$-feasible solution with probability at least $1-p$.
\end{theorem}

The following table summarizes the results for different numbers of buckets $b$.
For $b=1$, \cref{thm:rand} yields a $(1+\eps,8)$-feasible solution for $(1,1)$-feasible and $\eps$-compressed instances, yielding \cref{thm:main_rand} together with \cref{thm:reduction}.
\begin{table}[H]
	\centering
	\begin{tabular}{@{}rrrr@{}}\toprule
		$b$ & $\alpha$ (\cref{thm:det}) & $\alpha$ (\cref{thm:rand}) & $\beta$ \\\midrule
		1 & $2+\eps$ & $1+\eps$ & 8\\
		2 & $4+\eps$ & $2+\eps$ & 5.197\\
		3 & $6+\eps$ & $3+\eps$ & 4.234\\
		4 & $8+\eps$ & $4+\eps$ & 3.739\\\bottomrule
	\end{tabular}
	\caption{Approximation guarantees $(\alpha,\beta)$ of the algorithms in \cref{thm:det} and \cref{thm:rand} for different numbers of buckets $b$.}\label{tab:tradeoffs}
\end{table}

\subsection{Rounding fractional solutions}

Throughout the analysis, we assume $0<\eps<1$, and $\OPT$ denotes a fixed $(1,1)$-feasible solution to the given $\eps$-compressed instance of SNUkC. 
Note that we can assume w.l.o.g. that for each $\ell \in [L]$, $|\OPT_\ell| \leq m_\ell := \lfloor k_{\le \ell} \rfloor - \lfloor k_{\le \ell-1} \rfloor$, as moving a center to an earlier level only increases its radius. 

Let $h\in \{0,\ldots,L\}$ be some threshold to be specified later. We consider radii $r_\ell$ with $\ell \le h$ as large, and as small otherwise.

The idea is to guess points close to centers of an optimum solution with large radius.
By guessing, we mean enumerating several options and iterating in each of the resulting branches.
To bound the width and the depth of our enumeration approach, we crucially exploit that the cumulative budgets grow geometrically, thus bounding the number of centers in $\OPT$ with a large radius.

At each iteration of the enumeration, we try to fractionally cover the remaining
uncovered points using only balls of small radius.
If we almost manage to do so, we apply rounding approaches from \cite{NUkC,10.1145/3173046}, paired with our bucketing approach.
If not, we guess a new center with large radius, guided by an optimum dual solution.

To this end, let $U \subseteq V$ be the subset of vertices that is not covered by our partial solution guessed by the algorithm. Consider the following linear program that has a variable $x_{v,\ell}$ for each $v \in V$ and each $\ell \in \{h+1,\ldots,L\}$.
As stated above, we try to cover all points by only using balls of small radius.
\begin{equation}\label{eq:primal}
	\begin{aligned}
		\max \quad && \lambda\\
		\text{s.t. }&&\sum_{\ell>h}\sum_{v\in\Ball(u,r_\ell)}x_{v,\ell} &\ge \lambda && \forall \, u\in U,\\
		&&\sum_{v\in V}x_{v,\ell}&\le m_\ell && \forall \, \ell>h,\\
		&&x&\ge0.
	\end{aligned}
\end{equation}
Define $\lambda(U,h)$ as the optimum of this LP, with $\lambda(\emptyset,h):=+\infty$.

If $\lambda(U,h) \ge 1-\eps$, we can round the fractional solution by a controllable violation of the budget constraint and dilation of the radii.
To do so, we exploit an LP-aware reduction from NUkC to RMFC-T established by \textcite{NUkC}.
Unfortunately, this reduction loses a factor of $8$ on the dilation of the radii.
By using a bucketing-idea, we establish a new trade-off between the two approximation factors, enabling us to achieve approximations with dilation smaller than $8$. 
More precisely, we round each positive radius to a power $\mu^j$, where $\mu>1$ is a parameter and $j$ is an integer, while leaving zero radii unchanged.
Then, we merge levels with equal rounded radii and partition the resulting levels into $b$ groups such that any two positive radii in the same group differ by a factor of at least $\mu^b$.
We then assign each client to a group that supplies a sufficient fraction of its coverage and apply the tree LP-aware reduction to RMFC-T within each group.
This yields a smaller dilation factor at the cost of reduced coverage, establishing the trade-off between the two approximation factors.
Recall that we defined $\rho_b = \frac{2(b+1)^{1+\frac{1}{b}}}{b}$.
When we specify a set $U\subseteq V$ to be covered, an $(\alpha,\beta)$-feasible solution need only cover $U$,
and its centers may lie anywhere in $V$.

\begin{lemma}\label{lem:buckets}
	Let $0<\epsilon<1$ and let $(V,d,k,r)$ be an instance of SNUkC with $L$ levels and $k_1 \ge \frac{L}{\epsilon}$. Let $b \in \mathbb{Z}_{\ge 1}$, and let $U \subseteq V$.
	If $\lambda(U,0) \ge 1-\eps$, then we can efficiently determine a $\left(\frac{b +\epsilon}{1-\eps},\rho_b\right)$-feasible solution covering all points in $U$. 
\end{lemma}

The condition $k_1 \ge \frac{L}{\epsilon}$ allows us to absorb the additive budget violation resulting from rounding the fractional solution into the allowed multiplicative budget violation.
We prove this lemma in \cref{sec:rounding}.

As in \cite{blauth2025}, for a given $(1,1)$-feasible $\eps$-compressed instance of SNUkC, let $\hat h$ be the smallest index in $\{0,\ldots,L-1\}$ such that $k_{\hat h+1}\ge L/\eps$, and set $\hat h:=L$ if no such index exists.
Similarly, let $\check h$ be the smallest index in $\{0,\ldots,\hat h-1\}$ such that $k_{\check h+1}\ge\hat h/\eps$, and set $\check h:=\hat h$ if no such index exists.
In particular, $0\le\check h\le\hat h\le L$.
Note that $\hat h\le1+\left\lceil\log_{1+\eps}(L/\eps^2)\right\rceil$.
If $\hat h>0$, we also have $\check h\le1+\left\lceil\log_{1+\eps}(\hat h/\eps^2)\right\rceil$; if $\hat h=0$, then $\check h=0$.

For each nonempty interval $[\check h+1,\hat h]$ or $[\hat h+1,L]$, the cutoff definitions ensure that its first budget is at least its number of levels divided by $\eps$. 
These bounds let us use
\cref{lem:buckets} to round fractional solutions.
Extending these intervals would no longer guarantee the condition $k_1 \ge \frac{L}{\eps}$ in \cref{lem:buckets}. We therefore handle the initial levels by enumeration.

In our deterministic algorithm, we will choose the threshold index $h = \check{h}$, whereas in our randomized algorithm, we can afford to choose $h = \hat{h}$.
Together with \cref{lem:buckets}, this yields the following rounding guarantees.
For $h =\check{h} < \hat{h}$, we cannot apply \cref{lem:buckets} to the whole interval $[\check{h}+1,L]$ as the conditions in \cref{lem:buckets} may not be met. Instead, we apply \cref{lem:buckets} separately to both intervals $[\check{h}+1, \hat{h}]$ and $[\hat{h}+1,L]$. For this, we assign each point to the interval in which it is covered to an extent of at least $\frac{1-\eps}{2}$, thus losing a factor of two since we need to scale the solution in each interval to recover a coverage of $1-\eps$.
	
\begin{lemma}\label{lem:rounding}
	Let $1 >\epsilon > 0$, and let $(V,d,k,r)$ be an $\epsilon$-compressed instance of SNUkC with $L$ levels. Let $U \subseteq V$, and let $h \in \{\hat{h},\check{h}\}$.
	If $\lambda(U,h) \ge 1-\eps$,
	we can, for every $b \in \mathbb{Z}_{\ge 1}$, efficiently compute an $(\alpha,\rho_b)$-feasible solution $C'$ covering all points in $U$, only using centers $(v,\ell)$ with $\ell > h$, and
	\begin{itemize}
		\item $\alpha = \frac{b +\epsilon}{1-\eps}$ if $h = \hat{h}$, and
		\item $\alpha = 2\frac{b +\epsilon}{1-\eps}$ if $h = \check{h}<\hat h$.
	\end{itemize}
    Moreover, for each $h < \ell \le L$, we have $|C'_{\le\ell}|\le\alpha\sum_{i=h+1}^{\ell}m_i$.
\end{lemma}

We will prove this together with the previous \cref{lem:buckets} in \cref{sec:rounding}.

The choice of the cutoff determines a trade-off between the enumeration and the rounding step. Choosing $h=\hat h$ allows us to round the remaining fractional cover with a small budget violation, but requires more extensive enumeration. Choosing $h=\check h$ reduces the enumeration effort, at the expense of the additional factor two in the budget violation. We will show that the deterministic enumeration can handle the prefix up to $\check h$, while randomization allows us to use $\hat h$.

In either case, \cref{lem:rounding} provides a stopping criterion: as soon as the uncovered vertices $U$ satisfy $\lambda(U,h)\ge 1-\eps$, we can complete the cover by rounding the fractional solution. If this condition fails, we use the dual of the LP to guide the next step of the enumeration, as described below.

\subsection{A novel primal-dual based enumeration approach}

If in one iteration of our enumeration approach, we have $\lambda(U,h) < 1-\eps$, we want to guess points close to centers with large radius in our fixed $(1,1)$-feasible solution $\OPT$.
To this end, let us first define in which case we call our current solution \emph{consistent} with $\OPT$.
In the analysis, we will always follow an execution branch that stays consistent with $\OPT$.

A \emph{search state} is a set $C\subseteq V\times[h]$
satisfying $|C_\ell|\le m_\ell$ for every $\ell\in[h]$.
In particular, for each $\ell \in [h]$, we have
\[
    |C_{\le\ell}|
    \le \sum_{i=1}^{\ell}m_i
    =\lfloor k_{\le\ell}\rfloor
    \le k_{\le\ell}.
\]
For each pair $(v,\ell)\in C$, we open a ball of radius
$2r_\ell$ at $v$. Define
\begin{equation*}
    U_C:=
    V\setminus
    \bigcup_{(v,\ell)\in C}\Ball(v,2r_\ell)
\end{equation*}
to be the set of clients not yet covered by the search state.
\begin{definition}\label{def:consistency}
	A search state $C\subseteq V\times[h]$ is called \emph{consistent} with $\OPT$ if its pairs can be assigned injectively to pairs of $\OPT$, preserving levels, so that $(c,\ell) \in C$ is assigned to $(v,\ell) \in \OPT$ only if
	\begin{equation*}
		d(v,c)\le r_\ell.
	\end{equation*}
	A pair in \(\mathrm{OPT}_{\le h}\) is called \emph{matched} if it lies in the image of this assignment and \emph{unmatched} otherwise.
\end{definition}
Whenever a consistent search state is considered, fix one such injective assignment. 
Note that since we increased the balls in $C$ by a factor of $2$, $C$ covers all points that are also covered by the matched pairs in $\OPT$.
Hence a matched optimal ball has no client left in $U_C$.
Thus, if we allow a dilation factor of $2$, we do not need to identify centers of an optimum solution, but just points that are covered by centers in $\OPT_{\le h}$.

To bound our enumeration, our guesses are guided by an optimum solution to the dual linear program of \eqref{eq:primal}.
The dual of \eqref{eq:primal} is
\begin{equation}\label{eq:dual}
	\begin{aligned}
		\min &&\sum_{\ell>h}m_\ell p_\ell\\
		\text{s.t. }&&\sum_{u\in U\cap\Ball(v,r_\ell)}w_u&\le p_\ell &&\forall \, v\in V,\ \ell>h,\\
		&&\sum_{u\in U}w_u&=1,\\
		&&w&\ge0,\\
		&&p&\ge0.
	\end{aligned}
\end{equation}
Here, we have a dual variable $w_u$ for each coverage constraint, and a variable $p_\ell$ for each budget constraint.

The vector $w$ can be interpreted as a probability distribution on the remaining clients. We extend $w$ by zero outside $U$ and, for $S\subseteq V$, write $w(S)=\sum_{u\in S}w_u$ for its mass. 
For $\ell > h$, $p_\ell$ bounds the $w$-mass of \emph{every} possible ball of radius $r_\ell$, including balls used by $\OPT$. Therefore $m_\ell p_\ell$ bounds the mass covered by at most $m_\ell$ such balls. 
For nonempty $U$, the primal and the dual are always feasible.
Indeed, for the primal, $x \equiv 0$ together with $\lambda =0$ is feasible, whereas for the dual, any probability distribution on $U$ together with $p \equiv 1$ is feasible.
Thus, strong duality gives
\begin{equation*}\label{eq:dualfailure}
	\sum_{\ell>h}m_\ell p_\ell=\lambda(U,h)<1-\eps.
\end{equation*}
However, since $\sum_{u\in U}w_u=1$, this means that the weight covered by unmatched centers in $\OPT_{\le h}$ is at least $\eps$. 
We exploit this to guess one point $u \in U$ with $u \in \Ball(v,r_\ell)$ for some unmatched $(v,\ell) \in  \OPT_{\le h}$, and open a ball at $u$ of radius $2r_\ell$, thus fully covering $\Ball(v,r_\ell)$.

\begin{lemma}\label{lem:heavy}
	Suppose $C\subseteq V\times[h]$ is a search state consistent with $\OPT$. Assume $U_C\ne \emptyset$ and $\lambda(U_C,h)<1-\eps$, and let $(w,p)$ be an optimal dual solution. 
	Let $\OPT' \subseteq \OPT_{\le h}$ be the set of unmatched pairs. Then,
	\[
	w\left( \bigcup_{(v,\ell) \in \OPT'} \Ball(v,r_\ell) \right)>\eps.
	\]
	Consequently, for one center $(v,\ell) \in \OPT'$, we have $w(\Ball(v,r_\ell)) > \frac{\eps}{k_{\le h}}$.
\end{lemma}
\begin{proof}
	The assumptions imply $h\ge1$: if $h=0$, then $C=\emptyset$ and $U_C=V$, and the indicator vector of $\OPT$ is feasible for \eqref{eq:primal} with $\lambda=1$, a contradiction. Hence $k_{\le h}\ge k_1\ge1$.
	The total mass of balls in $\OPT_{>h}$ is at most $1-\eps$. Indeed,
	\[
	\sum_{(v,\ell)\in \OPT_{>h}}w(U_C\cap\Ball(v,r_\ell))
	\le\sum_{\ell>h}|\OPT_\ell|p_\ell
	\le\sum_{\ell>h}m_\ell p_\ell=\lambda(U_C,h)<1-\eps.
	\]
	Every remaining client $u \in U_C$ is covered in $\OPT$ either by a ball with small radius or by an unmatched ball in $\OPT_{\le h}$. Since $w(U_C)=1$, this proves 
	\[
	w\left( \bigcup_{(v,\ell) \in \OPT'} \Ball(v,r_\ell) \right)> \eps.
	\]
	Since there are at most $k_{\le h}$ centers in $\OPT'$, at least one of these balls has weight greater than $\frac{\eps}{k_{\le h}}$.
\end{proof}
The two conclusions in \cref{lem:heavy} serve different purposes: the deterministic search needs one heavy ball, whereas the randomized analysis needs the whole union to have mass greater than $\eps$.

\subsection{Randomized enumeration}

A straightforward idea is to sample, guided by an optimal dual solution, a center that we add to our search state. 

More precisely, given a search state $C \subseteq V \times [h]$ with $\lambda(U_C,h) < 1-\eps$, sample $u$ according to the weights $w$ given by an optimal dual solution, namely, for $v \in V$, $\Pr[u=v] = w_v$.
We add $(u,\ell)$ to $C$ by enumerating all possible choices of $\ell \in [h]$ and iterate.
This results in the following algorithm. 

For better readability, an \textbf{output} instruction terminates the entire algorithm. 
If the initial call finishes without output, the algorithm reports failure.

\begin{algorithm}[h]
		\caption{\textsc{RandomizedEnumeration}($C$)}
		\label{alg:random}
		\begin{algorithmic}[1]
				\State Let
				$U_C:=V\setminus
				\bigcup_{(v,\ell)\in C}\Ball(v,2r_\ell)$.
				
				\If{$U_C=\emptyset$}
				\State \textbf{output} $C$
				\EndIf
				
				\State Solve \eqref{eq:primal} for
				$U=U_C$.
				\If{$\lambda(U_C,h)\ge1-\eps$}
				\State Use \cref{lem:rounding} to compute
				a solution $C'$ covering $U_C$.
				\State \textbf{output} $C\cup C'$
				\EndIf
				
				\State Compute an optimal dual solution $(w,p)$
				for $U=U_C$.
				\State Sample $u\sim w$.
				
				\ForAll{$\ell\in[h]$ with $|C_\ell|<m_\ell$}
				\State
				\Call{RandomizedEnumeration}{$C\cup\{(u,\ell)\}$}
				\EndFor
			\end{algorithmic}
	\end{algorithm}

The high-level idea is that according to \cref{lem:heavy}, with probability at least $\eps$, we sample a vertex $u$ that is covered by an unmatched pair $(v,\ell) \in \OPT_{\le h}$. 
Note that by opening a ball of radius $2r_\ell$ at $u$, we fully cover $\Ball(v,r_\ell)$.
Thus, in each iteration, given a search state that is consistent with $\OPT$, one of the children remains consistent with probability at least $\eps$.
Since for $h = \hat{h}$, a branch has at most $k_{\le \hat{h}} = \mathcal{O}_\eps(L) = \mathcal{O}_\eps(\log n)$ extensions, we succeed with probability at least $n^{-\mathcal{O}_\eps(1)}$.

\begin{lemma}\label{lem:rand_correctness}
	Given a $(1,1)$-feasible $\eps$-compressed instance of SNUkC, \textsc{Randomized\-Enumeration}$(\emptyset)$ with $h = \hat{h}$ returns a $\left(\frac{b+\eps}{1-\eps}, \rho_b\right)$-feasible solution with probability at least $\eps^{\lfloor k_{\le\hat h}\rfloor} = n^{-\mathcal{O}_\eps(1)}$.
\end{lemma}

To analyze the running time of the algorithm, we observe that the budgets restrict the number of times each level can be chosen along a branch.
Writing $M:=\sum_{\ell=1}^{h}m_\ell=\lfloor k_{\le h}\rfloor$, the fully expanded tree of admissible level choices has
$\binom{M}{m_1,\ldots,m_{h}}$ leaves.
Indeed, this coefficient counts the ordered sequences of length $M$ in which each level $\ell$ occurs exactly $m_\ell$ times.
As every root-to-leaf path in the fully expanded tree has at most $M+1$ recursive calls (including the initial call), the number of recursive calls is at most
\[
    (M+1)\binom{M}{m_1,\ldots,m_{h}}.
\]
For $\eps$-compressed instances and $h = \hat{h}$, this is in $n^{\mathcal{O}_{\eps}(1)}$.

\begin{lemma}\label{lem:rand_running_time}
	Given a $(1,1)$-feasible $\eps$-compressed instance of SNUkC,
\textsc{Randomized\-Enumeration}$(\emptyset)$ with $h = \hat{h}$ runs in time
$n^{\mathcal{O}_{\eps}(1)}$.
\end{lemma}

We formally prove these lemmas in \cref{sec:algo_analysis}.
After adjusting the accuracy parameter and repeating the algorithm a polynomial number of times, they yield \cref{thm:rand}.
For completeness, we provide a formal proof in \cref{sec:parm_readjustment}. 

\subsection{Deterministic enumeration}

As an alternative to sampling a vertex $v$ according to the weights in an optimal solution, we can, if $\lambda(U_C,h) < 1-\eps$, in polynomial time compute a small set of points such that for one of these points $u$, we have $u \in \Ball(v,r_\ell)$ for some unmatched pair $(v,\ell) \in \OPT$, given that the current search state is consistent with $\OPT$.
Note that by opening a ball of radius $2r_\ell$ at $u$, we fully cover $\Ball(v,r_\ell)$.

\begin{lemma}\label{lem:det_enum}
	Let $h \ge 1$, and let $C\subseteq V\times[h]$ be a search state with $U_C\ne \emptyset$ and $\lambda(U_C,h)<1-\eps$.
	We can in polynomial time compute a set $H \subseteq U_C$ such that
	\begin{itemize}
		\item if $C$ is consistent with $\OPT$, then for some $u \in H$ there is an unmatched pair $(v,\ell) \in \OPT_{\le h}$ with $u \in \Ball(v,r_\ell)$, and
		\item $|H| \le 1+\ceil{\frac{k_{\le h}}{\eps}\ln\bigl(h|V|\bigr)}$.
	\end{itemize}
\end{lemma}

This follows from a standard greedy hitting-set
argument; see, e.g., \cite{Lovasz1975}.
For completeness, we provide a proof of the following lemma in \cref{sec:hitting-set}.

\begin{lemma}\label{lem:hitting}
	For a finite set $U$, let $w \in \mathbb{R}_{\ge 0}^U$ with $w(U)=1$.
	Let $\eps > 0$, and let $\emptyset\ne\mathcal{S} \subseteq 2^U$ with $w(S) \ge \eps$ for each $S \in \mathcal{S}$. In time polynomial in $|U|$ and $|\mathcal{S}|$, one can find a set $H \subseteq U$ with $H \cap S \neq \emptyset$ for each $S \in \mathcal{S}$, and
	\[
	|H| \le 1+\ceil{\frac{\ln |\mathcal{S}|}{\eps}}.
	\]
\end{lemma}

\begin{proof}[Proof of \cref{lem:det_enum}]
	Compute an optimal dual solution $(w,p)$ for $U=U_C$ and the given threshold $h$. Let
	\begin{equation*}\label{eq:heavyfamily}
		\mathcal{S}_C=\left\{U_C\cap\Ball(v,r_\ell):
		\begin{array}{l}
			v\in V,\ \ell\le h,
			w(U_C\cap\Ball(v,r_\ell))\ge \frac{\eps}{k_{\le h}}
		\end{array}\right\}.
	\end{equation*}
	If $\mathcal{S}_C=\emptyset$, set $H=\emptyset$.
	Otherwise, apply \cref{lem:hitting} with threshold
	$\eps/k_{\le h}$ to obtain a hitting set $H\subseteq U_C$.
	Since $|\mathcal{S}_C|\le h|V|$, in either case we have
	\[
	|H|\le 1+\ceil{\frac{k_{\le h}}{\eps}\ln(h|V|)}.
	\]
	
	Now suppose that $C$ is consistent with $\OPT$.
	By \cref{lem:heavy}, there is an unmatched pair
	$(v,\ell)\in\OPT_{\le h}$ such that
	$U_C\cap\Ball(v,r_\ell)\in\mathcal{S}_C$.
	Since $H$ hits this set, it contains a point
	$u\in\Ball(v,r_\ell)$, as required.
\end{proof}

Now we are ready to state our recursive deterministic enumeration approach.
We branch on the choice of a point $u$ in the hitting set from \cref{lem:det_enum} and a level $\ell\in[h]$ with $|C_\ell|<m_\ell$.
In each branch, we open a single ball of radius $2r_\ell$ at $u$ and recurse.
The recursion depth is at most $k_{\le h}$, as we open a new center in each iteration in which we do not terminate.

\begin{algorithm}[h]
    \caption{\textsc{DeterministicEnumeration}($C$)}
    \label{alg:det-short}
    \begin{algorithmic}[1]
        \State Let $U_C:=V\setminus\bigcup_{(v,\ell)\in C}\Ball(v,2r_\ell)$.
        \If{$U_C=\emptyset$}
            \State \textbf{output} $C$
        \EndIf
        \State Solve \eqref{eq:primal} for $U=U_C$.
        \If{$\lambda(U_C,h)\ge1-\eps$}
            \State Use \cref{lem:rounding} to compute a solution $C'$ covering $U_C$.
            \State \textbf{output} $C\cup C'$
        \EndIf
        \State Use \cref{lem:det_enum} to compute a small hitting set $H\subseteq U_C$.
        \ForAll{$(u,\ell)\in H\times[h]$ with $|C_\ell|<m_\ell$}
            \State \Call{DeterministicEnumeration}{$C\cup\{(u,\ell)\}$}
        \EndFor
    \end{algorithmic}
\end{algorithm}

\begin{lemma}\label{lem:det_correctness}
	Given a $(1,1)$-feasible $\eps$-compressed instance of SNUkC, \textsc{Deterministic\-Enumeration}$(\emptyset)$ for $h=\check{h}$ returns a $\left(2\frac{b+\eps}{1-\eps}, \rho_b\right)$-feasible solution.
\end{lemma}
\begin{lemma}\label{lem:det_running_time}
	Given a $(1,1)$-feasible $\eps$-compressed instance of SNUkC,
\textsc{Deterministic\-Enumeration}$(\emptyset)$ for $h=\check{h}$ runs in time
$n^{\mathcal{O}_{\eps}(1)}$.
\end{lemma}

We prove both lemmas in \cref{sec:algo_analysis}.
After adjusting the accuracy parameter, \cref{lem:det_correctness} and \cref{lem:det_running_time} yield \cref{thm:det}.
For completeness, we provide a formal proof in \cref{sec:parm_readjustment}.

\section{Analysis of the enumeration procedures}\label{sec:algo_analysis}

\subsection{Running time}

For the analysis of the running time, we need to analyze the number of execution branches we create. 
Each recursive call takes polynomial time as we can solve the primal LP \eqref{eq:primal} and the dual LP \eqref{eq:dual} efficiently, and the rounding and hitting-set constructions in \cref{lem:rounding,lem:det_enum} run in polynomial time as well.

We first bound the number of ways to choose the levels along a branch.

\begin{lemma}\label{lem:level_sequences}
    Let $h\in\{0,\ldots,L\}$ and consider an $\eps$-compressed instance.
    The number of sequences of levels in $[h]$ in which each level $i$
    occurs at most $m_i$ times, including the empty sequence, is at most
    \[
        (\lfloor k_{\le h}\rfloor+1)\,2^{(1+1/\eps)k_{\le h}}.
    \]
\end{lemma}
\begin{proof}
    For $h=0$, the empty sequence is the only choice. Suppose $h\ge1$
    and write $M_i:=\sum_{j=1}^i m_j=\lfloor k_{\le i}\rfloor$.
    First count the sequences in which every level $i$ occurs exactly $m_i$ times.
    Starting with the empty sequence, insert all occurrences of each level
    $i=1,\ldots,h$ in turn. At step $i$, we choose the $m_i$ positions of
    level $i$ among the $M_i$ positions in the resulting sequence.
    Thus the number of such sequences is
    \[
        \prod_{i=1}^h\binom{M_i}{m_i}
        \le 2^{\sum_{i=1}^h M_i}.
    \]
    Since $k_{\le i+1}\ge(1+\eps)k_{\le i}$, we have
    \[
        \sum_{i=1}^h M_i
        \le k_{\le h}\sum_{j=0}^{h-1}(1+\eps)^{-j}
        \le (1+1/\eps)k_{\le h}.
    \]
    Every valid sequence is a prefix of one of the sequences counted
    above, and each of those has $M_h+1$ prefixes.
\end{proof}

By the definitions of $\hat h$ and $\check h$, we have $k_\ell<L/\eps$ for $\ell\le\hat h$ and $k_\ell<\hat h/\eps$ for $\ell\le\check h$. Together with the geometric growth of the cumulative budgets, this yields
\[
k_{\le \hat{h}} \le \frac{(1+\eps)}{\eps^2} L,
\]
and 
\[
k_{\le \check{h}} \le \frac{(1+\eps)}{\eps^2} \hat{h} \le \frac{(1+\eps)}{\eps^2} \left(1+ \left\lceil \log_{1+\epsilon} \left( \frac{L}{\epsilon^2} \right) \right\rceil\right).
\]

Now, we are ready to prove \cref{lem:rand_running_time} and \cref{lem:det_running_time}.

\begin{proof}[Proof of \cref{lem:rand_running_time}]
    Each call samples at most one point and makes at most one recursive call
    per level. Thus each sequence of level choices gives rise to at most
    one call. For $h=\hat h$, we have $k_{\le h}=\mathcal{O}_\eps(L)
    =\mathcal{O}_\eps(\log(n+1))$. By \cref{lem:level_sequences}, the
    number of calls is $n^{\mathcal{O}_\eps(1)}$, and each takes polynomial time.
\end{proof}

\begin{proof}[Proof of \cref{lem:det_running_time}]
	As before, each call takes polynomial time.
    Let $h=\check h$. If $h=0$, there is only the initial call.
    Otherwise, by \cref{lem:det_enum}, each hitting set has size at most
    \[
        a:=1+\left\lceil\frac{k_{\le h}}{\eps}\ln(hn)\right\rceil.
    \]
    For a fixed sequence of levels of length $t$, there are at most $a^t$ calls,
    since at each step we choose one point from the current hitting set.
    As $t\le\lfloor k_{\le h}\rfloor$, it suffices to multiply the bound in
    \cref{lem:level_sequences} by $a^{k_{\le h}}$.
    The bound in \cref{lem:level_sequences} is clearly in $n^{\mathcal{O}_\eps(1)}$ for $h = \check{h}$, using $\check{h} \le \hat{h}$ and the argument in the proof of \cref{lem:rand_running_time}.
    Thus, it remains to bound $a^{k_{\le h}}$.

    For $h = \check{h}$, we have $k_{\le h}=\mathcal{O}_\eps(\log(L+1))
    =\mathcal{O}_\eps(\log\log(n+2))$. Since $h\le L$, the bound on $a$
    also gives $\log a=\mathcal{O}_\eps(\log\log(n+2))$.
    Thus
    \[
    \log(a^{k_{\le h}})
    = k_{\le h}\log a
    = \mathcal{O}_\eps\bigl((\log\log(n+2))^2\bigr)
    = \mathcal{O}_\eps(\log(n+1)).
    \]
    Hence $a^{k_{\le h}}=n^{\mathcal{O}_\eps(1)}$, as desired.
\end{proof}

\subsection{Correctness}

In this subsection, we prove \cref{lem:rand_correctness} and \cref{lem:det_correctness}.
For this, we first need to argue that both algorithms return a solution. 
To do so, we show that given a search state consistent with $\OPT$ (note that the initial search state $C = \emptyset$ is consistent with $\OPT$), we either have $\lambda(U_C,h) \ge 1-\eps$, in which case we output a solution, or among the execution branches we create there is one that remains consistent with $\OPT$. 
In our randomized enumeration, the latter will hold with constant probability. 

\begin{lemma}\label{lem:termination}
	Given a $(1,1)$-feasible $\eps$-compressed instance of SNUkC, \textsc{Deterministic\-Enumeration}$(\emptyset)$ always returns a solution. 
	Moreover, \textsc{Randomized\-Enumeration}$(\emptyset)$ returns a solution with probability at least $\eps^{\lfloor k_{\le\hat h}\rfloor}$.
\end{lemma}

During the proof, we follow a branch that remains consistent with $\OPT$.
Whenever the stopping condition fails, \cref{lem:heavy}
guarantees that the union of the balls of unmatched pairs in \(\OPT_{\le h}\) has dual weight greater than \(\eps\).
The deterministic hitting set contains a point in one of
the corresponding balls, while the randomized algorithm samples such
a point with probability at least $\eps$.
For this point, one of the choices of level preserves
consistency and matches one additional pair in $\OPT_{\le h}$.
Only a bounded number of such extensions are possible.
Once all pairs in $\OPT_{\le h}$ are matched, we will have $\lambda(U_C,h)\ge 1-\eps$, and the algorithm outputs a solution.

\begin{proof}
    Fix a search state $C$ that is consistent with $\OPT$,
    together with a witnessing injective assignment.
    If $U_C=\emptyset$, the algorithm returns $C$.
    If $\lambda(U_C,h)\ge1-\eps$, it completes the cover using
    \cref{lem:rounding}. It therefore remains to consider
    \[
        U_C\ne\emptyset,
        \qquad
        \lambda(U_C,h)<1-\eps.
    \]

    By \cref{lem:heavy}, the union of the balls of unmatched pairs in $\OPT_{\le h}$ has dual weight greater than $\eps$.
    If $u\in U_C$ belongs to one such ball $\Ball(v,r_\ell)$,
    then adding $(u,\ell)$ preserves consistency: assign this
    new pair to $(v,\ell)$ and retain all previous assignments.
    This extension is admissible because
    \[
        |C_\ell|<|\OPT_\ell|\le m_\ell.
    \]

    In the deterministic algorithm, \cref{lem:det_enum} ensures
    that the hitting set contains such a point $u$.
    Since the algorithm enumerates every admissible level,
    at least one child remains consistent with $\OPT$.

    In the randomized algorithm, the sampled point belongs to
    the union of the balls of unmatched pairs in $\OPT_{\le h}$ with probability
    at least $\eps$. Conditional on this event, enumerating the
    admissible levels again produces a consistent child.

    Every consistent extension matches one additional pair in $\OPT_{\le h}$. Once all these pairs are matched,
    every remaining client is covered by $\OPT_{>h}$, which
    gives a feasible solution to \eqref{eq:primal} with $\lambda\ge1$.
    Hence a consistent branch outputs a solution after
    at most
    \begin{equation}\label{eq:num_iterations}
        |\OPT_{\le h}|
        \le \lfloor k_{\le h} \rfloor 
    \end{equation}
    extensions.

    The deterministic search explores a consistent child whenever
    necessary and therefore always succeeds.
    In an iteration of the randomized search, a consistent branch stays consistent with probability at least $\eps$. Using \eqref{eq:num_iterations}, this gives a success probability of at least $\eps^{\lfloor k_{\le h} \rfloor }$.
\end{proof}

It remains to bound the quality of every solution returned by either algorithm.

The enumerated balls cover all points outside the remaining
set $U_C$, and the rounding step covers $U_C$.
To bound the number of balls, we use that enumeration respects
the budgets of the first $h$ levels, while rounding uses only
the budgets of the remaining levels.
Combining the two solutions therefore preserves the budget
factor of the rounding step.

\begin{lemma}\label{lem:sol_quality}
	Given a $(1,1)$-feasible $\eps$-compressed instance of SNUkC, a solution returned by \textsc{Randomized\-Enumeration}$(\emptyset)$ is $\left(\frac{b+\eps}{1-\eps},\rho_b\right)$-feasible.
	Moreover, a solution returned by \textsc{Deterministic\-Enumeration}$(\emptyset)$ is $\left(2\frac{b+\eps}{1-\eps},\rho_b\right)$-feasible.
\end{lemma}

Indeed, \cref{lem:rand_correctness,lem:det_correctness} directly follow from \cref{lem:termination,lem:sol_quality}.

\begin{proof}
    For the randomized algorithm, let
    \[
        h:=\hat h,
        \qquad
        \alpha:=\frac{b+\eps}{1-\eps},
    \]
    and for the deterministic algorithm, let
    \[
        h:=\check h,
        \qquad
        \alpha:=2\frac{b+\eps}{1-\eps}.
    \]
    Consider a call that outputs a solution, and let $C$ be
    its search state. By the conditions for adding a center, we have $|C_i|\le m_i$ for each $i \in [h]$.

    If $U_C=\emptyset$, the algorithm returns $C$.
    This set covers all points with dilation two and respects
    every cumulative budget, since
    \[
        |C_{\le\ell}|
        \le \sum_{i=1}^{\ell}m_i
        =\lfloor k_{\le\ell}\rfloor
        \le k_{\le\ell}
    \]
    for each $\ell \in [L]$.
    As $\alpha\ge1$ and $\rho_b\ge2$, the claim follows
    in this case.

    Otherwise, the algorithm returns $C\cup C'$, where $C'$
    is obtained from \cref{lem:rounding}.
    The balls in $C$ cover $V\setminus U_C$ with dilation two,
    and those in $C'$ cover $U_C$ with dilation $\rho_b$.
    Their union therefore covers $V$ with dilation $\rho_b$.

    It remains to check the cumulative budgets.
    If $h=0$, then $C=\emptyset$, so the required bound follows
    directly from \cref{lem:rounding}.

    Suppose now that $h>0$.
    By the guarantees in \cref{lem:rounding}, we have, for each $\ell > h$,
    \[
        |C'_{\le\ell}|
        \le \alpha\sum_{i=h+1}^{\ell}m_i.
    \]
    Moreover, $C'$ contains no centers at levels at most $h$.

    Hence, for $\ell\le h$, we have
    \[
        |(C\cup C')_{\le\ell}|
        =|C_{\le\ell}|
        \le \sum_{i=1}^{\ell}m_i
        =\lfloor k_{\le\ell}\rfloor
        \le \alpha k_{\le\ell}.
    \]
    For $\ell>h$, we obtain
    \[
        \begin{aligned}
        |(C\cup C')_{\le\ell}|
        &\le
        \sum_{i=1}^{h}m_i
        +\alpha\sum_{i=h+1}^{\ell}m_i\\
        &\le
        \alpha\sum_{i=1}^{\ell}m_i\\
        &=\alpha\lfloor k_{\le\ell}\rfloor\\
        &\le\alpha k_{\le\ell},
        \end{aligned}
    \]
    where the second inequality uses $\alpha\ge1$.
    This proves the claimed approximation guarantees.
\end{proof}

\section{Rounding fractional solutions}\label{sec:rounding}

In this section, we prove \cref{lem:buckets} and its implication \cref{lem:rounding}.
For the proof, we make use of the LP-aware reductions to RMFC-T by \textcite{NUkC}. 
For cleaner presentation and improved approximation guarantees, we reduce to the variant called 
Smooth Resource Minimization for Fire Containment on Trees introduced in \cite{blauth2025}. 

Before formally stating the reduction, we introduce some notation.
In \emph{Smooth Resource Minimization for Fire Containment on Trees} (SRMFC-T), we are given a rooted tree $G=(V,E)$ with root $r$ and depth $L$ (we call $L$ the number of levels), and budgets $k \in \mathbb{R}_{\ge 0}^L$.
Let $\Gamma \subseteq V$ denote its leaves.
For \(v\in V\), let \(P_v\) denote the unique path from \(r\) to \(v\), including \(v\).
The task is to compute a set $R \subseteq V \setminus \{r\}$ of vertices to fireproof, so that 
\begin{itemize}
	\item for each leaf $t \in \Gamma$, the unique $r$-$t$-path $P_t$ in $G$ contains some vertex in $R$, and
	\item $|R \cap V_{\le \ell}| \le \sum_{i=1}^\ell k_i$ for each $\ell \in [L]$, where $V_{\le \ell}$ denotes the vertices with distance at most $\ell$ to $r$ in $G$.
\end{itemize}
In the following, we leverage known sparseness results for the following polytope containing all $\alpha$-approximate fractional solutions.
For a set $S\subseteq V$, write
$x(S):=\sum_{v\in S}x_v$.
Every path $P_v$ includes its endpoint $v$.
We write

\[
Q^\RMFC_{\alpha}
:=
\left\{
x\in \mathbb{R}_{\ge 0}^V:
\begin{array}{ll}
x_r=0,\\[0.5ex]
x(V(P_t))\ge 1
& \forall \, t\in\Gamma,\\[0.5ex]
\displaystyle
\sum_{v\in V_{\le\ell}}x_v
\le\alpha k_{\le\ell}
& \forall \, \ell\in[L]
\end{array}
\right\}.
\] 
\textcite{10.1145/3173046} have shown that extreme points of this polytope are sparse, which we exploit to round fractional solutions as done in prior work. 
Let $x\in Q^\RMFC_{\alpha}$.
We call $v\in\supp(x)$ \emph{loose} if $x(V(P_v))<1$.
\textcite{10.1145/3173046} obtained the following bound on the number of loose vertices, assuming that $x$ is an extreme point solution. 
We again use the formulation from \cite{blauth2025Full}.

\begin{lemma}[\cite{10.1145/3173046}, direct implication of Lemma 9.1 in \cite{blauth2025Full}]\label{lem:RMFC_sparseness}
	Consider an instance of SRMFC-T with $L$ levels. 
	Let $\alpha \in \mathbb{R}_{\ge 0}$ and let $x$ be a vertex of $Q^\RMFC_{\alpha}$. 
	Then, at most $L$ support vertices of $x$ are loose.
\end{lemma}

For an extreme point $x\in Q^\RMFC_\alpha$, define
\[
R:=
\{v\in V:x_v \ge 1\}
\cup
\{v\in\supp(x):v\text{ is loose}\}.
\]
Since $x_r=0$, the root does not belong to $R$.

The set $R$ meets every root-to-leaf path.
Indeed, the first support vertex on such a path
either has value at least one or is loose.
Moreover, \cref{lem:RMFC_sparseness} gives
\[
|R\cap V_{\le\ell}|
\le
\sum_{v\in V_{\le\ell}}x_v+L
\le
\alpha k_{\le\ell}+L
\qquad \forall \, \ell\in[L].
\]
Thus, rounding increases every cumulative budget
by an additive term of at most $L$.

Next, we state the reduction to RMFC-T in \cite{NUkC}. We use a slight reformulation to fit our notation. 

For $\alpha,\beta\ge0$ and $U\subseteq V$, define
\[
Q_{\alpha,\beta}(U)
:=
\left\{
x\in \mathbb{R}_{\ge 0}^{V\times[L]}:
\begin{array}{ll}
\displaystyle
\sum_{\ell=1}^L
\sum_{v\in\Ball(u,\beta r_\ell)}x_{v,\ell}\ge1
& \forall \, u\in U,\\[1.2ex]
\displaystyle
\sum_{i=1}^{\ell}\sum_{v\in V}x_{v,i}
\le\alpha k_{\le\ell}
& \forall \, \ell\in[L]
\end{array}
\right\},
\]
containing all fractional $(\alpha,\beta)$-feasible solutions covering $U$.
Only the clients are restricted to $U$;
centers remain eligible at every point of $V$.

\begin{lemma}[Adapted from \cite{NUkC}]
\label{lem:reduction}
Let $(V,d,k,r)$ be an instance of SNUkC with $L$ levels,
and let $\emptyset\ne U\subseteq V$.
Suppose that
\[
r_\ell \ge\eta r_{\ell+1}
\qquad \forall \, \ell \in[L-1],
\]
where $\eta>1$.
Let $\alpha,\beta\ge0$ with
$Q_{\alpha,\beta}(U)\ne\emptyset$.

Then, we can efficiently construct an instance of
SRMFC-T with $L$ levels and budgets $k$ such that
$Q^\RMFC_\alpha\ne\emptyset$.
For $\ell \in [L]$, let $W_\ell$ denote the vertices at level $\ell$ of the
constructed tree.
Every integral tree cover $R$ can
be transformed efficiently into a cover of $U$ using
at most $|R\cap W_\ell|$ balls of radius
\[
\frac{2\eta}{\eta-1}\beta r_\ell
\]
at each level $\ell \in [L]$.
\end{lemma}

For completeness, we provide a proof of this version of the reduction
in \cref{app:metric-tree}, closely following the proof in \cite{NUkC}.

Now, we are ready to prove \cref{lem:buckets}. 
As noted earlier,  \cref{lem:buckets} establishes a new trade-off for rounding fractional solutions.

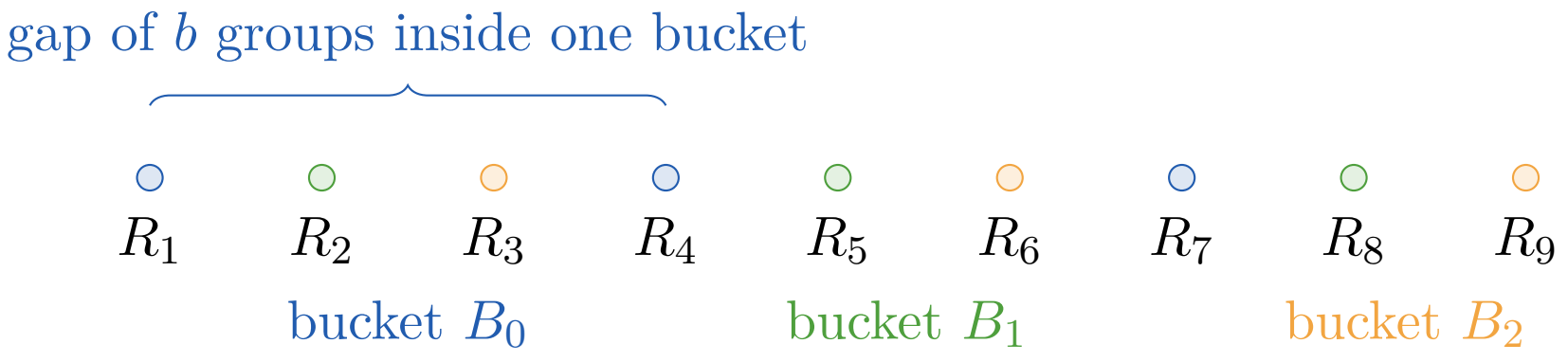
\begin{figure}[!t]
	\centering
	\begin{tikzpicture}[
		scale=1.2,
		every node/.style={font=\normalsize},
		radius/.style={circle,draw=black,fill=white,inner sep=1.8pt},
		bucketA/.style={circle,draw=ETHBlue,fill=ETHBlue!15,inner sep=1.8pt},
		bucketB/.style={circle,draw=ETHGreen,fill=ETHGreen!15,inner sep=1.8pt},
		bucketC/.style={circle,draw=ETHOrange,fill=ETHOrange!18,inner sep=1.8pt}
		]
		\foreach \i/\lab/\sty in {
			0/{\(R_1\)}/bucketA,
			1/{\(R_2\)}/bucketB,
			2/{\(R_3\)}/bucketC,
			3/{\(R_4\)}/bucketA,
			4/{\(R_5\)}/bucketB,
			5/{\(R_6\)}/bucketC,
			6/{\(R_7\)}/bucketA,
			7/{\(R_8\)}/bucketB,
			8/{\(R_9\)}/bucketC} {
			\node[\sty] at (\i+0.3,0) {};
			\node[below=4pt] at (\i+0.3,0) {\lab};
		}
		\node[ETHBlue] at (1.8,-0.85) {bucket \(B_0\)};
		\node[ETHGreen] at (4.7,-0.85) {bucket \(B_1\)};
		\node[ETHOrange] at (7.6,-0.85) {bucket \(B_2\)};
		\draw[decorate,decoration={brace,amplitude=4pt},ETHBlue] (0.3,0.42) -- (3.3,0.42)
		node[midway,above=5pt] {gap of \(b\) groups inside one bucket};
	\end{tikzpicture}
	\caption{Illustration of how we partition the groups of rounded radii into buckets for $b=3$. The group radii satisfy $R_j\ge\mu R_{j+1}$, so two positive radii in the same bucket differ by at least a factor of $\mu^b$.}
	\label{fig:bucketing-radii}
	
\end{figure}

\begin{proof}[Proof of \cref{lem:buckets}]
	If $U=\emptyset$, there is nothing to prove. Otherwise, choose a feasible solution $x$ of \eqref{eq:primal} with $h=0$ providing coverage at least $1-\eps$ to every vertex in $U$.
	Fix $\mu>1$ to be chosen later.
	We start by rounding up the positive radii to the next power of $\mu$, setting for each $\ell \in [L]$
	\[
		r'_\ell:=
		\begin{cases}
			\mu^{\lceil\log_\mu r_\ell\rceil} & \text{if } r_\ell>0,\\
			0 & \text{if } r_\ell=0.
		\end{cases}
	\]
	Thus $r_\ell\le r'_\ell\le\mu r_\ell$ for every $\ell\in[L]$.
	Merge consecutive levels with equal rounded radius into groups $G_1,\ldots,G_q$, ordered from largest to smallest radius, and put for each $j \in [q]$
	\[
		R_j:=r'_\ell\quad(\ell\in G_j),
		\qquad M_j:=\sum_{\ell\in G_j}m_\ell,
		\qquad z_{v,j}:=\sum_{\ell\in G_j}x_{v,\ell}.
	\]
	Then $\sum_{v\in V}z_{v,j}\le M_j$, and every vertex in $U$ still receives coverage at least $1-\eps$ from the rounded balls.

	Instead of directly applying \cref{lem:reduction} to the rounded instance, we sort the groups into $b$ \emph{buckets}
	\[
		B_i:=\{j\in[q]:(j-1)\bmod b=i\}
	\]
	for each $i=0,\ldots,b-1$.
	This way, we have $R_j\ge\mu^b R_{j'}$ for each pair $j<j'$ in the same bucket. Indeed, consecutive positive group radii differ by a factor of at least $\mu$, and the inequality also holds when $R_{j'}=0$.
	See \cref{fig:bucketing-radii} for an illustration.

	By the pigeonhole principle, we can partition $U$ into sets $U_0,\ldots,U_{b-1}$ so that for each $i=0,\ldots,b-1$ and each $u\in U_i$,
	\[
		\sum_{j\in B_i}\sum_{v\in\Ball(u,R_j)}z_{v,j}
		\ge\frac{1-\eps}{b}.
	\]
	Set \(a:=b/(1-\eps)\). Then \(az\) gives every vertex in \(U_i\) coverage at least one from pairs in bucket \(i\).
	While bucketing loses an additional factor of $b$ on the number of centers we open, the larger separation between radii in the same bucket reduces the loss in dilation in the LP-aware reduction to SRMFC-T.

    For $i=0,\ldots,b-1$, let $L_i:=|B_i|$.
	Fix a bucket $i$ with $U_i\ne\emptyset$ and consider the instance $(V,d,k(i),r(i))$ with $L_i$ levels and
	\[
		k(i):=(M_j)_{j\in B_i},\qquad r(i):=(R_j)_{j\in B_i},
	\]
	where the entries are listed in increasing order of $j$, and require coverage only of $U_i$.
	For this instance, the polytope $Q_{a,1}(U_i)$ is non-empty, as witnessed by the restriction of $az$ to this bucket.
	By \cref{lem:reduction} with $\eta=\mu^b$, we get an instance of SRMFC-T with $Q_a^\RMFC\ne\emptyset$.
	Choose an extreme point of this polytope, and select all vertices of value at least one and all loose vertices.
	By \cref{lem:RMFC_sparseness}, this gives an integral tree cover with an additional additive violation of each cumulative budget of at most $L_i$.
	The transformation in \cref{lem:reduction} does not increase the number of selected vertices at any level and produces a cover of $U_i$ using balls of radius
	\[
		\frac{2\mu^b}{\mu^b-1}R_j
	\]
	at group $j$. Buckets with $U_i=\emptyset$ require no balls.

	Taking the union of these solutions covers $U$.
	Assign each ball produced for group $G_j$ to the original level $\max G_j$, and let $A\subseteq V\times[L]$ be the resulting solution.
	Since $R_j\le\mu r_\ell$ for every $\ell\in G_j$, the dilation with respect to the original radii is at most
	\[
		\frac{2\mu^{b+1}}{\mu^b-1}.
	\]
	It remains to analyze the budget violation.
	For any original prefix $\ell \in [L]$, only groups with $\max G_j\le\ell$ contribute balls. 
	Hence
	\[
	\begin{aligned}
		|A_{\le\ell}|
		&\le a\sum_{j:\,\max G_j\le\ell}M_j+\sum_{i=0}^{b-1}L_i\\
		&\le a\sum_{j=1}^{\ell}m_j+L\\
		&=a\lfloor k_{\le\ell}\rfloor+L\\
		&\le(a+\eps)k_{\le\ell}\\
		&\le\frac{b+\eps}{1-\eps}k_{\le\ell},
	\end{aligned}
	\]
	where we used $k_{\le\ell}\ge k_1\ge L/\eps$ to absorb the additive term $L$.

    Choosing $\mu=(b+1)^{1/b}$ gives
    \[
        \frac{2\mu^{b+1}}{\mu^b-1}
        =\frac{2(b+1)^{1+1/b}}{b}=\rho_b,
    \]
    and proves the desired guarantee.
    For fixed $b$, comparisons with powers of $\mu$ can be
    performed exactly by taking $b$th powers, since
    $\mu^b=b+1$.
\end{proof}

With this, we are now ready to prove \cref{lem:rounding}.

\begin{proof}
	If $U=\emptyset$, return the empty set. Otherwise, let $x$ be a feasible solution providing coverage at least $1-\eps$ to every vertex in $U$.
	
	If $h=\hat h=0$, then $k_1\ge L/\eps$, so we directly apply \cref{lem:buckets} to the original instance. If $h=\hat h>0$, then $h<L$ because $U\ne\emptyset$ and $\lambda(U,h)>0$. 
	In this case, we apply \cref{lem:buckets} to the instance $(V,d,m',r')$, where $m'$ and $r'$ arise from $m$ and $r$ by deleting the first $h$ entries. Its first budget satisfies
	\[
		m'_1 = m_{h+1}\ge k_{h+1}-1
		\ge \frac{L}{\eps}-1
		\ge \frac{L-h}{\eps}.
	\]
	
	It remains to consider the case $h=\check h<\hat h$.
	This implies $\hat h < L$, as otherwise $\check h = \hat h = L$.
	We split the remaining levels into two instances $(V,d,m^{(1)},r^{(1)})$ and $(V,d,m^{(2)},r^{(2)})$. The vectors $m^{(1)}$ and $r^{(1)}$ contain the entries of $m$ and $r$ from index $\check h+1$ to $\hat h$, and $m^{(2)}$ and $r^{(2)}$ arise from $m$ and $r$ by deleting the first $\hat h$ entries.
	Let $U_1\subseteq U$ consist of all vertices $u\in U$ with
	\[
		\sum_{\ell = \check{h}+1}^{\hat{h}}
		\sum_{v\in\Ball(u,r_\ell)}x_{v,\ell}
		\ge \frac{1-\eps}{2},
	\]
	and let $U_2:=U\setminus U_1$. Every vertex in $U_2$ receives coverage at least $(1-\eps)/2$ from levels greater than $\hat h$.
	
	Now, on both instances, scale the restriction of $x$ by a factor of $2$, and also increase the budgets by a factor of $2$. Each resulting fractional solution provides coverage at least $1-\eps$ to the vertices assigned to its instance. 
	Note that
	\[
	2m^{(1)}_1 = 2m_{\check{h}+1} \ge 2\lfloor k_{\check{h}+1} \rfloor \ge k_{\check{h}+1} \ge \frac{\hat{h}}{\eps} \ge \frac{\hat{h} - \check{h}}{\eps}
	\]
	and 
	\[
	m^{(2)}_1 \ge m_{\hat{h}+1} \ge k_{\hat{h}+1} -1 \ge \frac{L}{\eps} -1 \ge \frac{L - \hat{h}}{\eps}
	\]
	by the definitions of the thresholds. Instances with no assigned clients are skipped. Applying \cref{lem:buckets} to each remaining instance and taking the union of the resulting covers yields the desired integral solution. 
	We remark that before taking the union, we restore the original level indices, mapping local level $j$ to $\check h+j$ in the first instance and to $\hat h+j$ in the second. 
	The dilation remains $\rho_b$, and the budget factor increases by a factor of $2$.
	
	For $h=\hat h=0$, the proof of \cref{lem:buckets} gives $|C'_{\le\ell}|\le \frac{b}{1-\eps}\lfloor k_{\le\ell}\rfloor+L$ for each $\ell \in [L]$. Since $\lfloor k_{\le\ell}\rfloor\ge L/\eps-1\ge (1-\eps)L/\eps$, this is at most $\alpha\lfloor k_{\le\ell}\rfloor$. In all other cases, the cumulative bounds within the intervals give
    \[
        |C'_{\le\ell}|\le\alpha\sum_{i=h+1}^{\ell}m_i
        \qquad \forall \, \ell>h.
    \]
    In both cases, the cumulative bounds imply the claimed bounds for the original instance, since
	\[
		\sum_{i\le \ell}m_i
		=\lfloor k_{\le \ell}\rfloor
		\le k_{\le \ell}
		\qquad \forall \, \ell \in[L].
	\]
\end{proof}

\section*{Origin of the results and use of AI.}
This work grew out of the second author's bachelor's thesis~\cite{PrussnerThesis}, which contains the radius-bucketing argument yielding an $(O_\varepsilon(\log^*n),2+\varepsilon)$-approximation, as well as a deterministic $(O_\varepsilon(1),4+\varepsilon)$-approximation based on a different enumeration approach.
The dual-witness search presented in this paper was subsequently proposed by GPT-6 Astra, which also supplied initial proofs of the resulting improvements.
GPT-6 Astra was also used to help prepare and revise parts of the manuscript.
The authors take responsibility for the mathematical results and their presentation.

\begingroup
\setlength{\emergencystretch}{1em}
\printbibliography
\endgroup

\newpage

\appendix
\crefalias{section}{appendix}

\section{Parameter readjustment}\label{sec:parm_readjustment}

\begin{proof}[Proof of \cref{thm:det,thm:rand}]
	Fix $b\in\mathbb{Z}_{>0}$ and $\eps>0$, and choose a positive
	rational number
	\[
	\eta\le\frac{\eps}{2b+2+\eps}.
	\]
	An $\eps$-compressed instance is also $\eta$-compressed.
	Run the algorithms with $\eta$ in place of $\eps$, including
	in the definitions of $\hat h$ and $\check h$.
	By \cref{lem:sol_quality}, the randomized and deterministic budget factors are at most
	\[
	\frac{b+\eta}{1-\eta}\le b+\frac{\eps}{2}
	\qquad\text{and}\qquad
	2\frac{b+\eta}{1-\eta}\le2b+\eps,
	\]
	respectively, and their dilation is $\rho_b$.
	The deterministic algorithm always succeeds by
	\cref{lem:termination}, proving \cref{thm:det} together with
	\cref{lem:det_running_time}.
	
	For the randomized algorithm, the proof of
	\cref{lem:termination} gives success probability at least
	\[
	s:=\eta^{\lfloor k_{\le\hat h}\rfloor}
	\]
	in one run. Since $k_{\le\hat h}=\mathcal{O}_\eta(\log(n+1))$,
	we have $s^{-1}=n^{\mathcal{O}_\eta(1)}$.
	We assume that, after polynomial-time preprocessing of the dual weights
	at each search state, we can draw an exact sample in $\mathcal{O}(1)$ time.
	
	Repeat the algorithm independently
	\[
	N:=\left\lceil s^{-1}\ln(1/p)\right\rceil
	\]
	times and return the first solution found.
	The probability that all runs fail is at most
	\[
	(1-s)^N\le e^{-Ns}\le p.
	\]
	Since $s^{-1}$ is polynomially bounded, \cref{lem:rand_running_time}
	implies that the total running time is polynomial in the input size
	and $\log(1/p)$ for fixed $b$ and $\eps$.
	This proves \cref{thm:rand}.
\end{proof}

\begin{proof}[Proof of \cref{thm:main_rand,thm:main}]
	For \cref{thm:main_rand}, choose a positive rational
	$\delta\le\min\{1/2,\eps/8\}$.
	Apply \cref{thm:rand} with $b=1$ and accuracy $\delta$,
	and then \cref{thm:reduction} with the same accuracy.
	The resulting factors satisfy
	\[
	(1+\delta)^2\le1+\eps,
	\qquad
	8(1+\delta)\le8+\eps,
	\]
	as desired.
	
	For \cref{thm:main}, observe that
	\[
	\rho_b=2(1+1/b)(b+1)^{1/b}\longrightarrow2
	\]
	as $b \to \infty$.
	Choose $b$ so that $\rho_b\le2+\eps/2$, and choose a
	positive rational
	$\delta\le\min\{1/2,\eps/(4+\eps)\}$.
	Applying \cref{thm:det} and \cref{thm:reduction} with
	accuracy $\delta$ gives
	\[
	(1+\delta)(2b+\delta)=\mathcal{O}_\eps(1),
	\qquad
	(1+\delta)\rho_b\le2+\eps.
	\]
	Both algorithms run in polynomial time for fixed $\eps$.
\end{proof}

\section{Approximation hardness}\label{sec:hardness}

In this section, we prove \cref{thm:hardness} using a simple
reduction from \emph{Minimum Dominating Set}.
Given a graph $G=(V,E)$, this problem asks for a minimum
cardinality set $D\subseteq V$ such that every vertex of $G$
either belongs to $D$ or has a neighbor in $D$.
We use the following hardness result.

\begin{theorem}[Direct consequence of \cite{DS14}]\label{thm:dominating_set_hardness}
	For every fixed $0<\eps<1$, it is NP-hard to approximate
	Minimum Dominating Set within a factor of $(1-\eps)\ln n$, where $n$ denotes the number of vertices in the given instance.
\end{theorem}

\begin{proof}[Proof of \cref{thm:hardness}]
	Fix $0<\eps<1$.
	Let $G=(V,E)$ be an instance of Minimum Dominating Set on $n=|V|\ge 1$ vertices.
	Define a metric $d$ on $V$ by
	$$
	d(u,v)=
	\begin{cases}
		1 & \text{if } \{u,v\}\in E,\\
		2 & \text{otherwise}
	\end{cases}
	$$
	for $u \neq v \in V$.
	The triangle inequality holds because all distances
	belong to $\{1,2\}$.
	For $h\in[n]$, consider the SNUkC instance
	$(V,d,k,r)$ with $r=\{1\}$ and $k=\{h\}$.
	
	A dominating set of size at most $h$ yields a feasible
	solution by placing a ball of radius one at each of its
	vertices.
	Conversely, the centers of any collection of balls of
	radius less than two covering $V$ form a dominating set.
	Indeed, each vertex is either a center or adjacent to a center.
	
	Suppose there exists a polynomial-time
	$((1-\eps)\ln n,2-\eps)$-approximation algorithm for SNUkC.
	Run it on the constructed SNUkC instance for every $h\in[n]$,
	retain each returned center set only if it dominates $G$, and return a smallest retained set.
	
	Let $\OPT_{\mathrm{DS}}$ be the minimum cardinality of a
	dominating set in $G$.
	For $h=\OPT_{\mathrm{DS}}$, the constructed SNUkC instance is
	feasible, so the algorithm returns at most
	$(1-\eps)\ln n\cdot\OPT_{\mathrm{DS}}$ centers covering
	$V$ at radius at most $2-\eps$.
	These centers form a dominating set.
	Thus we obtain a polynomial-time
	$((1-\eps)\ln n)$-approximation for Minimum Dominating Set, contradicting
	\cref{thm:dominating_set_hardness}.
\end{proof}

\section{Proof of the greedy hitting set argument}\label{sec:hitting-set}

\begin{proof}[Proof of \cref{lem:hitting}]
	Start with $H = \emptyset$.
	Let $\mathcal{S}' \subseteq \mathcal{S}$ denote the collection of sets $S$ with $S \cap H = \emptyset$.
	As long as $\mathcal{S}' \neq \emptyset$, weighted averaging gives
	\[
	\sum_{u\in U}w_u\,|\{S\in\mathcal{S}':u\in S\}|
	=\sum_{S\in\mathcal{S}'}w(S)\ge\eps|\mathcal{S}'|.
	\]
	Since the coefficients $w_u$ are nonnegative and sum to one, there is some point $u\in U$ with 
	\[
	|\{S\in\mathcal{S}':u\in S\}|\ge\eps|\mathcal{S}'|.
	\]
	Thus some point belongs to at least an $\eps$ fraction of the remaining sets. Greedily choose a point $u$ belonging to the most remaining sets, add $u$ to $H$, and delete those sets $S \in \mathcal{S}'$ with $u \in S$. After $t$ iterations at most $|\mathcal{S}|(1-\eps)^t\le |\mathcal{S}|e^{-\eps t}$ sets remain. After $1+\ceil{\frac{\ln |\mathcal{S}|}{\eps}}$ iterations, this upper bound is strictly less than one. Since the number of remaining sets is an integer, no set remains.
\end{proof}

\section{The metric-to-tree reduction}
\label{app:metric-tree}

We construct the tree by clustering clients from small
radii to large radii.
The choice of representatives preserves fractional
coverage, while separation bounds the distance from
a selected tree vertex to all clients below it.

\begin{proof}[Proof of \cref{lem:reduction}]
	Choose $x\in Q_{\alpha,\beta}(U)$.
	For $p\in U$ and $\ell\in[L]$, define
	\[
	c_\ell(p):=
	\sum_{u\in\Ball(p,\beta r_\ell)}x_{u,\ell},
	\qquad
	c_{\ge \ell}(p):=
	\sum_{i=\ell}^L c_i(p).
	\]
	These sums include centers throughout $V$.
	
	We first construct level $L$.
	Choose an unassigned client $p\in U$, create a tree
	vertex represented by $p$, and assign to it all
	still-unassigned clients $q\in U$ with
	$d(p,q)\le2\beta r_L$.
	Repeat until every client is assigned.
	The resulting vertices are the leaves of the tree;
	their assigned clients are not additional
	tree vertices.
	
	Suppose that level $\ell+1$ has been constructed.
	Among its still-unassigned vertices, choose one whose
	representative $p$ minimizes $c_{\ge \ell+1}(p)$.
	Create a level-$\ell$ vertex represented by $p$, and
	attach every still-unassigned level-$(\ell+1)$ vertex
	represented by $q$ with $d(p,q)\le2\beta r_\ell$.
	Repeat until every level-$(\ell+1)$ vertex is assigned.
	After constructing level one, attach its vertices
	to an artificial root $r$.
	
	Write $W_\ell$ for the vertices at level $\ell$ and
	$W_{\le\ell}:=\bigcup_{i=1}^{\ell}W_i$.
	For a level-$\ell$ vertex $v$ represented by $p$, put
	\[
	y_v^0:=c_\ell(p),
	\quad \text{and} \quad
	y_v:=\min\{1,y_v^0\}.
	\]
	Set $y_r^0=y_r:=0$.
	We verify the budget constraints, path constraints,
	and lifting property.
	
	First, distinct representatives selected at level $\ell$
	are at distance greater than $2\beta r_\ell$.
	Their radius-$\beta r_\ell$ neighborhoods in $V$ are
	therefore pairwise disjoint.
	Consequently,
	\[
	\sum_{v\in W_\ell}y_v
	\le
	\sum_{v\in W_\ell}y_v^0
	\le
	\sum_{u\in V}x_{u,\ell}.
	\]
	Summing over the first $\ell$ levels gives
	\[
	\sum_{v\in W_{\le\ell}}y_v
	\le
	\sum_{i=1}^{\ell}\sum_{u\in V}x_{u,i}
	\le
	\alpha k_{\le\ell}.
	\]
	
	Next, before truncation, every path from a level-$\ell$
	vertex represented by $p$ to a leaf has value at least
	$c_{\ge \ell}(p)$.
	We prove this by induction from level $L$ upwards.
	At level $L$, the assertion is immediate.
	For the induction step, let $q$ represent the child
	on the path.
	It was available when $p$ was chosen, so
	\[
	c_{\ge \ell+1}(q)\ge c_{\ge \ell+1}(p).
	\]
	By induction, the child-to-leaf path has value at least
	$c_{\ge \ell+1}(q)$.
	Adding the value $c_\ell(p)$ at the parent proves the claim.
	
	At level one, $c_{\ge1}(p)\ge1$ because $p\in U$.
	Thus every root-to-leaf path has total $y^0$-value
	at least one.
	Truncation preserves this: a path containing a truncated
	coordinate retains coverage at least one, and every other path
	is unchanged.
	Together with the budget bounds, this proves
	$y\in Q^\RMFC_\alpha$.
	
	It remains to transform an integral tree cover.
	Let a selected level-$\ell$ vertex be represented by $p$,
	and let $u$ be a client assigned to a leaf below it.
	Consecutive representatives on the tree path have
	distance at most $2\beta r_i$ when the parent is at
	level $i$.
	The final step from the leaf representative to $u$
	has distance at most $2\beta r_L$.
	Hence
	\[
	d(p,u)
	\le
	2\beta\sum_{i=\ell}^L r_i
	\le
	2\beta r_\ell\sum_{j=0}^{\infty}\eta^{-j}
	=
	\frac{2\eta}{\eta-1}\beta r_\ell.
	\]
	Open a ball of this radius at every selected
	representative.
	Every client is assigned to a leaf whose path meets
	the tree cover, so the resulting balls cover $U$.
	Each selected vertex produces one ball at the same
	level, and deleting duplicates can only decrease
	the counts.
	
	There are at most $|U|$ vertices at each level,
	and all construction steps take polynomial time.
\end{proof}
\end{document}